\documentclass[prd,twocolumn,amsmath,amssymb]{revtex4}
\usepackage{graphicx}
\begin{document}
\title{\bf \boldmath Experimental studies of the $\pi^+\pi^-\pi^+\pi^-\pi^0$,
$K^+K^-\pi^+\pi^-\pi^0$ and $p\bar p\pi^+\pi^-\pi^0$ final states
produced in $e^+e^-$ annihilation at $\sqrt{s}=$ 3.773 and 3.650
GeV}
\author{
\vspace{0.35cm}
{\large BES Collaboration}\\
\vspace{0.35cm} M.~Ablikim$^{1}$,              J.~Z.~Bai$^{1}$,
Y.~Ban$^{12}$, X.~Cai$^{1}$,                  H.~F.~Chen$^{16}$,
H.~S.~Chen$^{1}$,              H.~X.~Chen$^{1}$, J.~C.~Chen$^{1}$,
Jin~Chen$^{1}$,                Y.~B.~Chen$^{1}$
% ...
\\
\vspace{0.15cm}
{\it
$^{1}$ Institute of High Energy Physics, Beijing 100049, People's Republic of China\\
% ...
$^{12}$ Peking University, Beijing 100871, People's Republic of China\\
% ...
$^{16}$ University of Science and Technology of China, Hefei 230026, People's Republic of China\\
% ...
}}

\begin{abstract}
We report measurements of the observed
cross sections for $e^+e^-\to\omega \pi^+\pi^-$, $\omega K^+K^-$, $\omega p\bar p$,
$K^+K^-\rho^0\pi^0$, $K^+K^-\rho^+\pi^-+c.c.$,
$K^{*0}K^-\pi^+\pi^0+c.c.$, $K^{*+}K^-\pi^+\pi^-+c.c.$,
$\phi\pi^+\pi^-\pi^0$ and $\Lambda \bar \Lambda \pi^0$ at $\sqrt s=$ 3.773 and 3.650 GeV.
Upper limits (90\% C.L.) are given for observed cross sections and for $\psi(3770)$
decay branching fractions for production of these final states.
These measurements are made by analyzing the data sets of 17.3 pb$^{-1}$ collected at
$\sqrt{s}=3.773$ GeV and 6.5 pb$^{-1}$ collected at $\sqrt{s}=3.650$ GeV with
the BES-II detector at the BEPC collider.
\end{abstract}

\maketitle

\oddsidemargin  -0.2cm \evensidemargin -0.2cm

\section{Introduction}
During the past thirty years,
the $\psi(3770)$ resonance was expected to decay almost entirely
into pure $D\bar D$ \cite{delco}.
However, previous data suggest that the $\psi(3770)$ is not saturated by
$D\bar D$ decays \cite{rzhc}.
Recently, the BES Collaboration measured the branching fraction for
$\psi(3770)\to$ non-$D\bar D$ decay to be about $(15\pm5)$\%
\cite{brdd1,brdd2,pdg07}
with two different data samples and analysis methods.
This indicates that the $\psi(3770)$
may substantially decay into charmless final states.
In the last two years, many efforts have been undertaken by the BES
\cite{bai,ablikim,kskl,rhopi,crshads,crshads2} and CLEO
\cite{adam,adams,coans,huang,cronin,briere} collaborations to search for
charmless decays of $\psi(3770)$ but only upper limits were derived
for most decay modes.
So far, the results can not explain the discrepancy between the observed
cross sections for $D\bar D$ and $\psi(3770)$ production.
To understand this discrepancy, a comparison of observed cross sections for exclusive
light-hadron final states at the center-of-mass energies of 3.773 GeV and
below 3.660 GeV excluding the contributions from $J/\psi$, $\psi(3686)$ and
$D\bar D$ decays could be helpful.
A measurement of exclusive cross sections at these two or at more energy
points can also provide valuable information to understand the mechanism of
the continuum light-hadron production.

In this paper, we report measurements of the observed
cross sections for the exclusive light-hadron final states,
of $\omega\pi^+\pi^-$, $\omega K^+K^-$, $\omega p\bar p$,
$K^+K^-\rho^0\pi^0$, $K^+K^-\rho^+\pi^-+c.c.$,
$K^{*0}K^-\pi^+\pi^0+c.c.$, $K^{*+}K^-\pi^+\pi^-+c.c.$,
$\phi\pi^+\pi^-\pi^0$ and $\Lambda \bar \Lambda \pi^0$, produced in
$e^+e^-$ annihilation, and derive upper limits for $\psi(3770)$ decays into
these final states.
The data sets used in the analysis were collected at $\sqrt{s}=$ 3.773 and 3.650 GeV
with the BES-II detector at the
BEPC collider, which correspond to the integrated luminosities of
17.3 pb$^{-1}$ and 6.5 pb$^{-1}$,
respectively. For convenience, we call these two data
sets the $\psi(3770)$ resonance data and the continuum data in the
paper, respectively.

\section{BES-II detector}

The BES-II is a conventional cylindrical magnetic detector that is
described in detail in Refs. \cite {bes,bes2}. A 12-layer Vertex
Chamber(VC) surrounding a beryllium beam pipe provides input to
event trigger, as well as coordinate information. A forty-layer
main drift chamber (MDC) located just outside the VC yields
precise measurements of charged particle trajectories with a solid
angle coverage of $85\%$ of 4$\pi$; it also provides ionization
energy loss ($dE/dx$) measurements for particle identification.
Momentum resolution of $1.7\%\sqrt{1+p^2}$ ($p$ in GeV/$c$) and
$dE/dx$ resolution of $8.5\%$ for Bhabha scattering electrons are
obtained for the data taken at $\sqrt{s}=3.773$ GeV. An array of
48 scintillation counters surrounding the MDC measures time of
flight (TOF) of charged particles with a resolution of about 180
ps for electrons. Outside the TOF counters, a 12 radiation length,
lead-gas, six-readout-layer barrel shower counter (BSC), operating
in limited streamer mode, measures the energies of electrons and
photons over $80\%$ of the total solid angle with an energy
resolution of $\sigma_E/E=0.22 /\sqrt{E}$ ($E$ in GeV) and spatial
resolutions of $\sigma_{\phi}=7.9$ mrad and $\sigma_Z=2.3$ cm for
electrons. A solenoidal magnet outside the BSC provides a 0.4 T
magnetic field in the central tracking region of the detector.
Three double-layer muon counters instrument the magnet flux return
and serve to identify muons with momentum greater than 500 MeV/c,
with a solid angle coverage of $68\%$.

\section{Event Selection}
\label{evtsel}

In selection of the above processes, the possible intermediate
resonances are searched by analyzing the final states
$\pi^+\pi^-\pi^+\pi^-\gamma\gamma$,
$K^+K^-\pi^+\pi^-\gamma\gamma$ and $p\bar
p\pi^+\pi^-\gamma\gamma$.
The $\pi^0$, $\omega$,
$\rho^0$, $\rho^+$, $K^{*0}$, $K^{*+}$, $\phi$ and $\Lambda$
particles are reconstructed by the decays $\pi^0\to\gamma\gamma$,
$\omega \to \pi^+\pi^-\pi^0$, $\rho^0\to\pi^+\pi^-$,
$\rho^+\to\pi^+\pi^0$, $K^{*0}\to K^{+}\pi^{-}$, $K^{*+}\to
K^+\pi^0$, $\phi\to K^+K^-$ and $\Lambda\to p\pi^-$, respectively.
Throughout this paper, charge conjugation is implied.

To select the candidate events, we require the number of charged tracks
to be four with total charge zero. Each charged track should be well
reconstructed in the MDC with good helix fits and satisfy
$|\rm{cos}\theta|<0.85$, where $\theta$ is the polar angle.
All tracks, save those from $\Lambda$ decays, must originate from the
interaction region by requiring that the closest approaches of a charged track
are less than 2.0 cm in the $xy$-plane and 20.0 cm in the $z$ direction.

The TOF and $dE/dx$ measurements for each charged track are used to
calculate the confidence levels $CL_\pi$, $CL_K$ or $CL_p$ for the
pion, kaon or proton hypotheses.
The pion candidate is required to have a confidence level $CL_\pi$
greater than 0.1\%. In order to reduce the misidentification,
the kaon candidate is required to have the confidence level $CL_K$
greater than $CL_\pi$. For proton identification, the ratio $\frac{CL_p}
{CL_\pi+CL_K+CL_p}$ is required to be greater than 0.6.

The BSC measurements are used to select photons. The energy of
each good photon deposited in the BSC should be greater than 0.05
GeV, and the electromagnetic shower should start in the first five
readout layers. The angle between the cluster development
direction and the photon emission direction is required to be
within 37$^\circ$ \cite{besdphy}. In order to reduce the radiative
photons from charge particles, the angle between the photon and
the nearest charged track is greater than 22$^{\circ}$ \cite{besdphy}.

In order to improve mass resolution and suppress combinatorial
background, an energy-momentum conservation kinematic fit is
imposed on the $\pi^+\pi^-\pi^+\pi^-\gamma\gamma$,
$K^+K^-\pi^+\pi^-\gamma\gamma$ or $p\bar p\pi^+\pi^-\gamma\gamma$
combination. In addition, we constrain the invariant mass of the
two photons to the $\pi^0$ nominal mass. The candidates with a
kinematic fit probability greater than $1\%$ are accepted.
If more than one combination remains after the above selection criteria,
the combination with the largest fit probability is retained.

For the $K^+K^-\pi^+\pi^-\pi^0$ final state, we exclude the events
from $D \bar D$ decays by rejecting the events in which the $D^0$
and $\bar D^0$ mesons can be reconstructed in the decay modes of
$D^0\to K^-\pi^+$ and $\bar D^0\to K^+\pi^-\pi^0$ \cite{besdcrs}.

\section{Data Analysis}

We search for possible intermediate resonances by examining the
invariant mass spectra of the $\pi^+\pi^-\pi^0$, $\pi^+\pi^-$,
$\pi^\pm\pi^0$, $K^{\pm}\pi^{\mp}$, $K^\pm\pi^0$, $K^+K^-$ and
$p\pi^-/\bar p\pi^+$ combinations from the selected $\pi^+\pi^-
\pi^+\pi^-\pi^0$, $K^+K^-\pi^+\pi^-\pi^0$ and $p\bar
p\pi^+\pi^-\pi^0$ events. The invariant masses of these
combinations are calculated with the fitted momentum vectors from
the kinematic fit. In the paper, they are denoted by
$M_{\pi^+\pi^-\pi^0}$, $M_{\pi^+\pi^-}$, $M_{\pi^\pm\pi^0}$,
$M_{K^{\pm}\pi^{\mp}}$, $M_{K^\pm\pi^0}$, $M_{K^+K^-}$ and $M_{p
\pi^-/\bar p\pi^+}$, respectively.

\subsection{Candidates for $e^+e^- \to \omega\pi^+\pi^-$,
$e^+e^- \to \omega K^+K^-$ and $e^+e^- \to \omega p\bar p$}
\begin{figure}[htbp]
\begin{center}
\includegraphics*[width=8.0cm]
{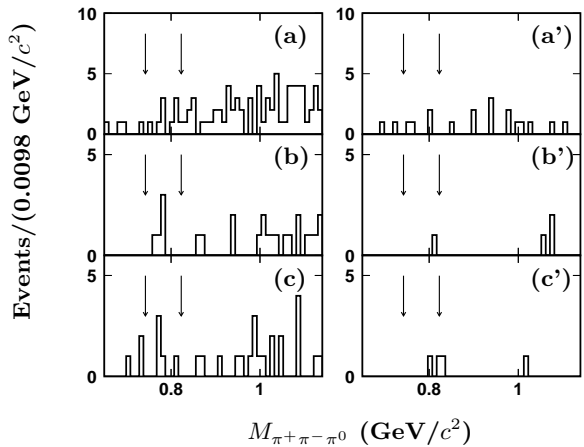} \put(-145,0){\bf
$M_{\pi^+\pi^-\pi^0}$ (GeV/$c^2$)}
\put(-235,45){\rotatebox{90}{\bf Events/(0.0098 GeV/$c^2$)}}
\put(-135.5,150){\bf (a)} \put(-135.5,104){\bf (b)}
\put(-135.5,58){\bf (c)} \put(-38,150){\bf (a')} \put(-38,104){\bf
(b')} \put(-38,58){\bf (c')} \caption{ The distributions of the
invariant masses of the $\pi^+\pi^-\pi^0$ combinations from the
selected (a) $\pi^+\pi^- \pi^+\pi^-\pi^0$, (b)
$K^+K^-\pi^+\pi^-\pi^0$ and (c) $p\bar p\pi^+\pi^-\pi^0$ events
from the $\psi(3770)$ resonance data (left) and the continuum data
(right).} \label{fig:xm3pi}
\end{center}
\end{figure}

To investigate the processes $e^+e^-\to \omega\pi^+\pi^-$, $e^+e^-\to
\omega K^+K^-$, and $e^+e^-\to \omega p\bar{p}$, we analyze the invariant
masses of the $\pi^+\pi^-\pi^0$ combinations from the
selected $\pi^+\pi^- \pi^+\pi^-\pi^0$, $K^+K^-\pi^+\pi^-\pi^0$ and
$p\bar p\pi^+\pi^-\pi^0$ events.
Figure \ref{fig:xm3pi} shows the invariant mass distributions from the
selected final states.
The mass window of $\pm$40 MeV/c$^2$
(accounting for the $\omega$ width \cite{pdg} and its mass resolution
determined by Monte Carlo simulation)
around the $\omega$ nominal mass is taken as the $\omega$ signal
region. Counting the events
with $M_{\pi^+\pi^-\pi^0}$ in the $\omega$ signal regions, we
obtain the numbers of the events in the signal region for searching for
the signal events $e^+e^- \to
\omega\pi^+\pi^-$, $e^+e^- \to \omega K^+K^-$ and $e^+e^- \to
\omega p\bar p$ observed from the $\psi(3770)$ resonance data (left) and
the continuum data (right), respectively. The numbers of the events observed
in the $\omega$ signal regions are listed in the second columns in tables \ref{tab:crs3773}
and \ref{tab:crs3650}.

\subsection{Further analyses of the $K^+K^-\pi^+\pi^-\pi^0$ final state}

\begin{figure}[htbp]
\begin{center}
\includegraphics*[width=8.0cm]
{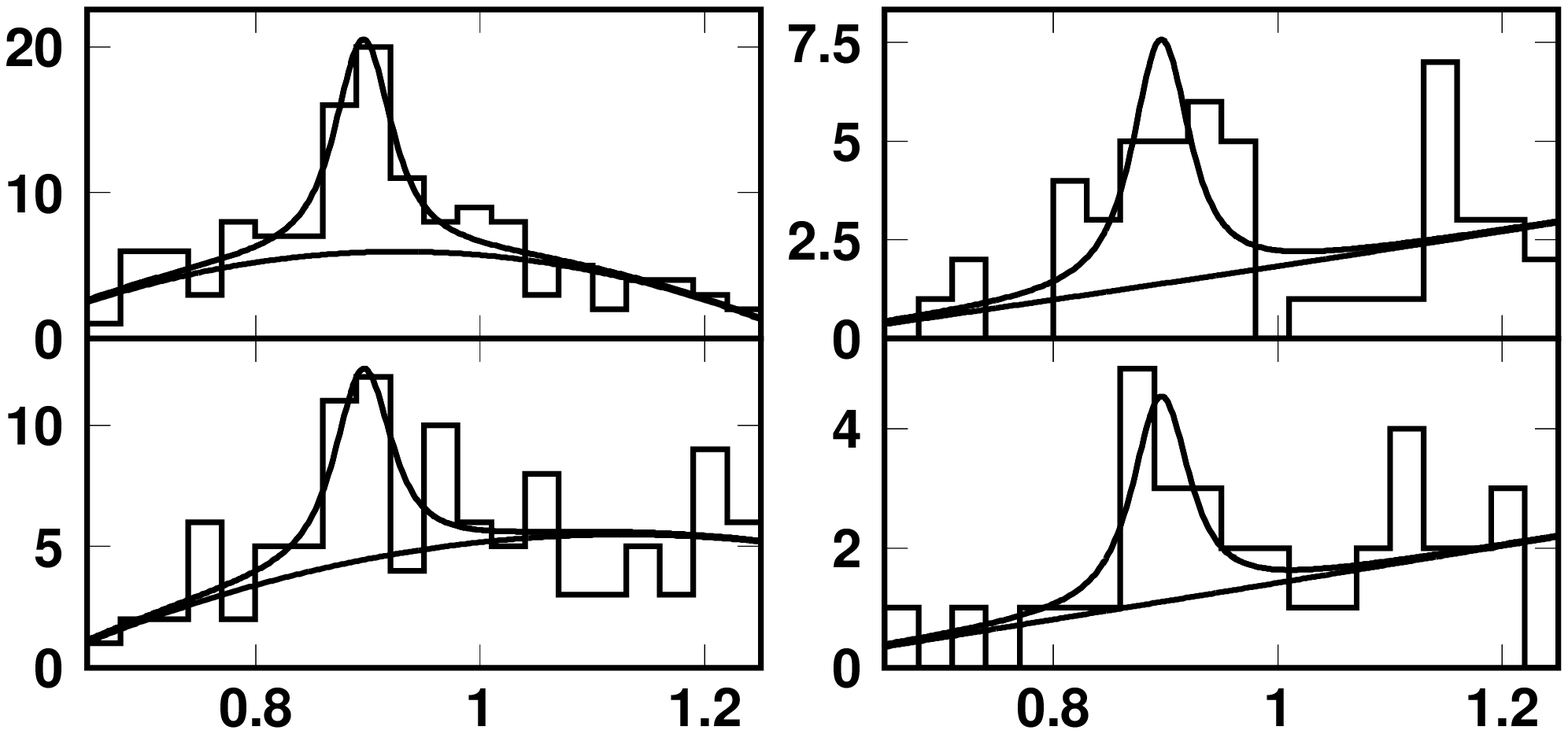} \put(-145,0){\bf $M_{K\pi}$
(GeV/$c^2$)} \put(-235,15){\rotatebox{90}{\bf Events/(0.02
GeV/$c^2$)}} \put(-195.5,95){\bf (a)} \put(-195.5,52){\bf (b)}
\put(-98,95){\bf (a')} \put(-98,52){\bf (b')} \caption{ The
distributions of the invariant masses for the (a) $K^\pm\pi^\mp$
and (b) $K^\pm\pi^0$ combinations from the selected
$K^+K^-\pi^+\pi^-\pi^0$ events from the $\psi(3770)$ resonance
data (left) and the continuum data (right).} \label{fig:xmkpi}
\end{center}
\end{figure}

Figure \ref{fig:xmkpi} shows the invariant masses of the
$K^{\pm}\pi^{\mp}$ and $K^{\pm}\pi^0$ combinations from the
selected $K^+K^-\pi^+\pi^-\pi^0$ events. In each figure, the
$K^{*}$ signal is clearly observed. Fitting to these invariant
mass spectra with a Breit-Wigner convoluted with a Gaussian
resolution function for the $K^*$ signal and a second polynomial
for the background, we obtain the numbers $N^{\rm obs}$ of the
signal events for $e^+e^- \to K^{*0}K^-\pi^+\pi^0+c.c.$ and
$e^+e^- \to K^{*+}K^-\pi^+\pi^-+c.c.$ observed from the
$\psi(3770)$ resonance data (left) and the continuum data (right).
These numbers are summarized in the second columns of
tables \ref{tab:crs3773} and \ref{tab:crs3650}. In the fits,
the mass and width of $K^{*}$ are fixed to the PDG values \cite{pdg}.

\begin{figure}[htbp]
\begin{center}
\includegraphics*[width=8.0cm]
{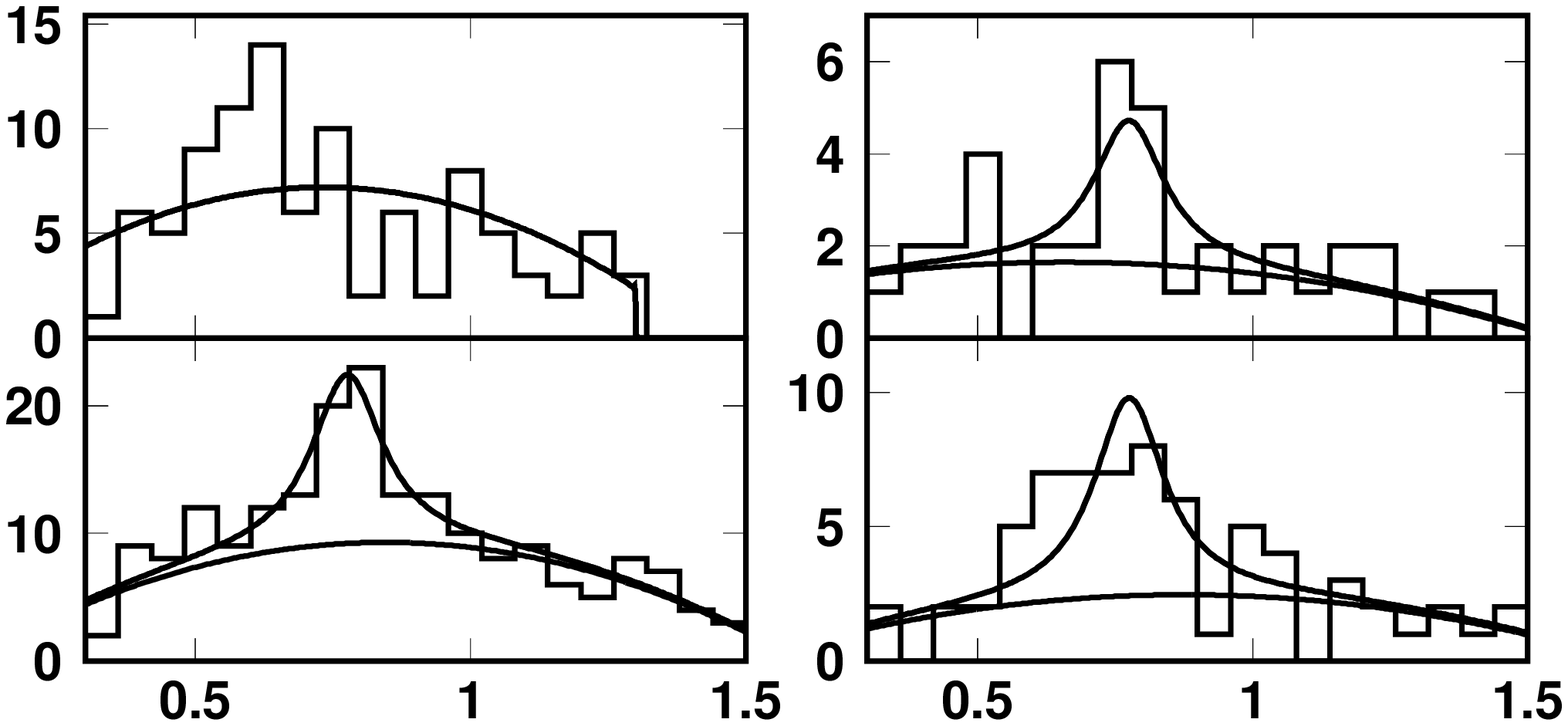}
\put(-145,0){\bf $M_{\pi\pi}$ (GeV/$c^2$)}
\put(-235,15){\rotatebox{90}{\bf Events/(0.06 GeV/$c^2$)}}
\put(-135.5,95){\bf (a)}
\put(-135.5,52){\bf (b)}
\put(-38,95){\bf (a')}
\put(-38,52){\bf (b')}
\caption{
The distributions of the invariant masses of the (a) $\pi^+\pi^-$
and (b) $\pi^{\pm}\pi^0$ combinations from the
selected $K^+K^-\pi^+\pi^-\pi^0$ events from the
$\psi(3770)$ resonance data
(left) and the continuum data (right).}
\label{fig:xmpipi}
\end{center}
\end{figure}

Figure \ref{fig:xmpipi} shows the invariant mass spectra of the
$\pi^+\pi^-$ and $\pi^{\pm}\pi^0$ combinations from the selected
$K^+K^-\pi^+\pi^-\pi^0$ events. Fitting to these invariant mass
spectra with a Breit-Wigner convoluted with a Gaussian resolution
function for the $\rho$ signal and a second order polynomial for
the background, we obtain the numbers $N^{\rm obs}$ of the signal
events for $e^+e^- \to K^+K^-\rho^0\pi^0$ and $e^+e^- \to
K^+K^-\rho^+\pi^-+c.c.$ from the $\psi(3770)$ data (left) and
the continuum data (right), which are shown in the second columns
of tables \ref{tab:crs3773} and \ref{tab:crs3650}.
In the fits, the mass and
width of $\rho$ are fixed to the PDG values \cite{pdg}.

\begin{figure}[htbp]
\begin{center}
\includegraphics*[width=8.0cm]
{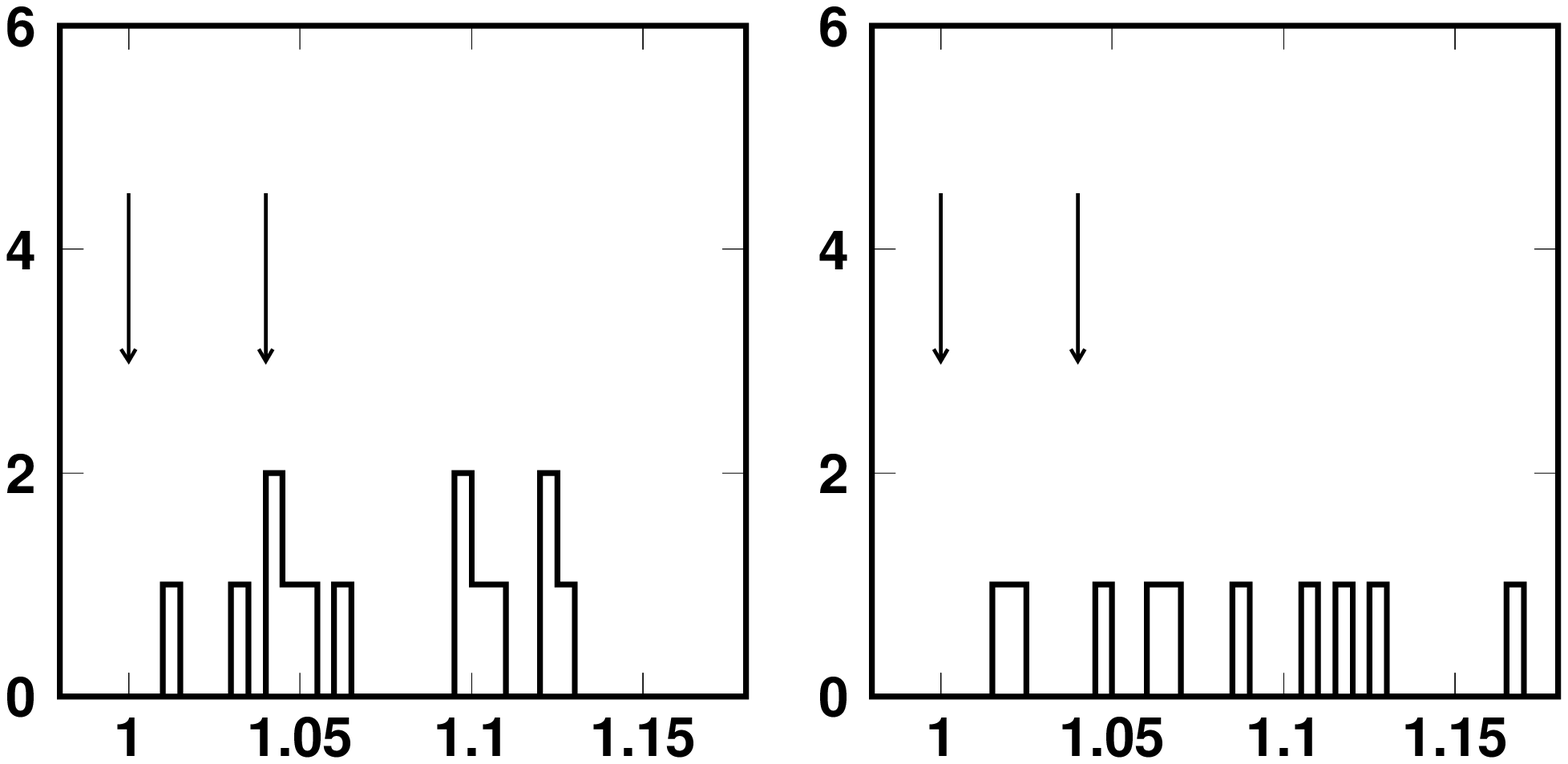}
\put(-145,0){\bf $M_{K^+K^-}$ (GeV/$c^2$)}
\put(-235,10){\rotatebox{90}{\bf Events/(0.005 GeV/$c^2$)}}
\caption{
The distributions of the invariant masses of the $K^+K^-$ combinations from
the selected $K^+K^-\pi^+\pi^-\pi^0$ events
from the $\psi(3770)$ resonance data (left) and the continuum data
(right), where the pairs of arrows represent the $\phi$ signal regions.}
\label{fig:xmkk}
\end{center}
\end{figure}

Figure \ref{fig:xmkk} shows the distributions of the invariant
masses of the $K^+K^-$ combinations from the selected
$K^+K^-\pi^+\pi^-\pi^0$ events. The mass window of $\pm$20 MeV/$c^2$
around the $\phi$ nominal mass is taken as the $\phi$ signal
region \cite{crshads}. Selecting the events with $M_{K^+K^-}$ in
the $\phi$ signal regions, we obtain 2 events in the signal region for
searching for the $\phi \pi^+\pi^-\pi^0$ final state observed from both
the $\psi(3770)$ resonance data (left) and the continuum data (right).

\subsection{Further analyses of the $p\bar p\pi^+\pi^-\pi^0$ final state}

\begin{figure}[htbp]
\begin{center}
\includegraphics*[width=8.0cm]
{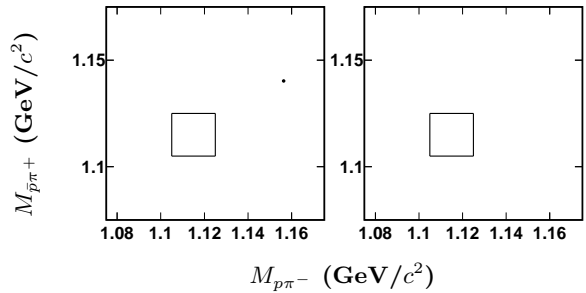} \put(-145,0){\bf $M_{p
\pi^-}$ (GeV/$c^2$)} \put(-235,25){\rotatebox{90}{\bf $M_{\bar p
\pi^+}$ (GeV/$c^2$)}} \caption{ The scatter plots of $M_{p \pi^-}$
versus $M_{\bar p \pi^+}$ from the selected $p\bar p
\pi^+\pi^-\pi^0$ events from the $\psi(3770)$ resonance data
(left) and the continuum data (right), where the rectangle regions
show the $\Lambda$ and $\bar \Lambda$ signal regions.}
\label{fig:xmppi_xmppi}
\end{center}
\end{figure}

The scatter plots of $M_{p\pi^-}$ versus $M_{\bar p \pi^+}$ from the selected $p\bar
p\pi^+\pi^-\pi^0$ events are shown in Fig. \ref{fig:xmppi_xmppi}.
The mass window of $\pm$10 MeV/$c^2$ around
the $\Lambda$ nominal mass is taken as the $\Lambda$ signal
region, determined by Monte Carlo simulation. In the $\Lambda \bar
\Lambda\pi^0$ signal region in each figure, no signal event for
the $\Lambda \bar \Lambda\pi^0$ final state is observed from the
two data sets.

\section{Background Subtraction}

For the selected candidate events, there are the
contributions from $J/\psi$ and $\psi(3686)$ decays due to ISR
(Initial State Radiation)
process. In addition, there are the contaminations from the other
final states due to misidentification between charged pions and
kaons, or due to missing photon(s), etc.. Above the $D\bar D$
threshold, there are also the contributions from $D\bar D$ decays.
The number $N^{\rm b}$ of these contributions should be subtracted from the
number $N^{\rm obs}$ of the selected events. These can be estimated by Monte Carlo
simulation, which has been discribed in Ref. \cite{crshads} in
detail.

In the following analyses, we ignore the interference effects
between the continuum and resonance amplitudes, since we don't
know the details about the two amplitudes. In this case,
subtracting $N^{\rm b}$ from $N^{\rm obs}$, the numbers $N^{\rm
net}$ of the signal events for the final states
$e^+e^-\to K^+K^-\rho^0\pi^0$, $e^+e^-\to K^+K^-\rho^+\pi^-+c.c.$,
$e^+e^-\to K^{*0}K^-\pi^+\pi^0+c.c.$ and
$e^+e^-\to K^{*+}K^-\pi^+\pi^-+c.c.$ are obtained.

For the other final states, only a few events are observed from
the data sets. We set the upper limits $N^{\rm up}$ on these
numbers of the signal events by using the Feldman-Cousins method
\cite{felman} in the absence of background at 90\% confidence
level (C.L.). These numbers are listed in tables \ref{tab:crs3773}
and \ref{tab:crs3650}.

\section{Results}

\subsection{Monte Carlo efficiency}

The detection efficiencies for reconstruction of the events of
$e^+e^-\to$ {\it exclusive light hadrons} are estimated by Monte Carlo
simulation with a phase space generator for the BES-II detector
\cite{bessim}, including initial state radiation and photon vacuum
polarization corrections \cite{isr} with $1/s$ cross section
energy dependence. Final state radiation \cite{fsr} decreases the
detection efficiency by not more than 0.5\%. Detailed Monte Carlo
analysis gives the detection efficiency for each final state at
$\sqrt{s}=$ 3.773 and 3.650 GeV. They are summarized in tables
\ref{tab:crs3773} and \ref{tab:crs3650}. For the final states
containing $\pi^0$, $\omega$, $\rho^0$, $\rho^+$, $K^{*0}$,
$K^{*+}$, $\phi$ and $\Lambda$ particles, the branching fractions for
the decays $\pi^0\to\gamma\gamma$, $\omega\to\pi^+\pi^-\pi^0$,
$\rho^0\to \pi^+\pi^-$, $\rho^+\to \pi^+\pi^0$, $K^{*0}\to
K^+\pi^-$, $K^{*+}\to K^+\pi^0$, $\phi\to K^+K^-$ and $\Lambda\to
p\pi^-$ are set to be 100\% in the generator. They are corrected with their
branching fractions quoted from PDG \cite{pdg}, see Eq. (\ref{eq:crs}) and
Eq. (\ref{eq:crsup}).

\subsection{Systematic error}
\label{sys}

The systematic errors in the measurement of the observed cross
section for $e^+e^-\to$ {\it exclusive light hadrons} arise mainly from
the uncertainties in luminosity ($\sim$2.1\% \cite{brdd1,brdd2}),
photon selection ($\sim$2.0\% per photon), tracking efficiency
($\sim$2.0\% per track), charged particle identification
($\sim$0.5\% per pion or kaon, $\sim$2.0\% per proton), kinematic
fit ($\sim$1.5\%), Monte Carlo statistics
[$\sim$(1.2$\sim$3.9)\%], branching fractions quoted from PDG
\cite{pdg} [$\sim$0.03\% for $\mathcal B(\pi^0\to\gamma\gamma)$,
$\sim$0.79\% for $\mathcal B(\omega\to \pi^+\pi^-\pi^0)$,
$\sim$1.22\% for $\mathcal B(\phi\to K^+K^-)$ and $\sim$0.78\% for
$\mathcal B(\Lambda\to p\pi^-)$], background subtraction
[$\sim$(0.0$\sim$13.4)\%], fit to mass spectrum
[$\sim$(1.9$\sim$16.5)\%], and Monte Carlo modeling ($\sim$6.0\%
\cite{crshads}). The total systematic errors $\Delta_{\rm sys}$
for each final state at $\sqrt{s}=$ 3.773 and 3.650 GeV are
obtained by adding these uncertainties in quadrature,
respectively, which are shown in tables \ref{tab:crs3773} and
\ref{tab:crs3650}.

\subsection{Observed cross section or its upper limit for $e^+e^-\to f$}

The observed cross section for $e^+e^-\to f$ is determined by
\begin{equation}
\sigma_{e^+e^-\to f} = \frac{N^{\rm net}} {\mathcal{L} \times
\epsilon [\times \prod_i^n B_i]}, \label{eq:crs}
\end{equation}
where $\mathcal{L}$ is the integrated luminosity of the data set,
$N^{\rm net}$ is the number of the signal events for $e^+e^- \to
f$, $\epsilon$ is the detection efficiency for the final state.
Here, $n=$ 1 or 2 or 3, is the number of the intermediate
resonances in the final state, $B_i$ denotes the branching
fraction \cite{pdg} for the intermediate resonance decay,
such as ${\mathcal B}(\pi^0 \to\gamma\gamma)$, ${\mathcal B}(\omega\to\pi^+\pi^-\pi^0)$,
${\mathcal B}(\phi\to K^+K^-)$ and ${\mathcal B}(\Lambda\to
p\pi^-)$ etc.. Inserting the corresponding numbers in Eq.
(\ref{eq:crs}), we obtain the observed cross sections for the
final states $e^+e^-\to K^+K^-\rho^0\pi^0$,
$e^+e^-\to K^+K^-\rho^+\pi^-+c.c.$, $e^+e^-\to K^{*0}K^-\pi^+\pi^0+c.c.$
and $e^+e^-\to K^{*+}K^-\pi^+\pi^-+c.c.$ at $\sqrt{s}=$ 3.773 and 3.650
GeV, respectively. They are summarized in tables \ref{tab:crs3773} and
\ref{tab:crs3650}, where the first error is statistical and the
second systematic.

For the other final states for which only a few events are observed from
the data, the upper limits on their observed cross section are set
by
\begin{eqnarray}
\sigma^{\rm up}_{e^+e^-\to f}= \frac{N^{\rm up}} {\mathcal{L}
\times \epsilon \times (1-\Delta_{\rm sys}) [\times \prod_i^n B_i]},
\label{eq:crsup}
\end{eqnarray}
where $N^{\rm up}$ is the upper limit on the number of the signal
events for $e^+e^- \to f$, and $\Delta_{\rm sys}$ is the
systematic error in the measurement of the observed cross section.
Inserting the corresponding numbers in Eq. (\ref{eq:crsup}), we
obtain the upper limits on the observed cross sections for these
final states at $\sqrt{s}=$ 3.773 and 3.650 GeV, respectively, which are
shown in tables \ref{tab:crs3773} and \ref{tab:crs3650}.

\begin{table*}[hbtp]
\begin{center}
\caption{The observed cross sections for $e^+e^-\to f$ at
$\sqrt{s}=$ 3.773 GeV, where $N^{\rm obs}$ is the number of events
observed from the $\psi(3770)$ resonance data, $N^{\rm b}$ is the
number of total background events, $N^{\rm net}$ is the number of
the signal events, $N^{\rm up}$ is the upper limit on the number
of the signal events, $\epsilon$ is the detection efficiency,
$\Delta_{\rm sys}$ is the relative systematic error in the
measurement, $\sigma$ is the observed cross section and
$\sigma^{\rm up}$ is the upper limit on the observed cross section
at 90\% C.L..}
\begin{tabular}{|l|c|c|c|c|c|c|} \hline
\multicolumn{1}{|c|}{$e^+e^-\to$}&$N^{\rm obs}$&$N^{\rm b}$ &$N^{\rm net}$ (or $N^{\rm
up}$) &$\epsilon$[\%]&$\Delta_{\rm sys}$ &$\sigma$ (or
$\sigma^{\rm up}$) [pb] \\ \hline
$\omega\pi^+\pi^-$        & 9            &  0         &$<15.30$      &$ 3.06\pm0.08$&0.116&$<37.1$  \\
$\omega K^+K^-$           & 5            &  0         &$<9.99$       &$ 1.67\pm0.06$&0.118&$<44.5$              \\
$\omega p\bar p$          & 5            &  0         &$<9.99$       &$ 3.69\pm0.09$&0.124&$<20.3$              \\
$K^+K^-\rho^0\pi^0$       &0             &  0         &$<2.44$       &$ 2.90\pm0.05$&0.114&$<5.6$               \\
$K^+K^-\rho^+\pi^-+c.c.$  &$ 48.6\pm15.4$&$ 2.7\pm0.9$&$ 45.9\pm15.4$&$ 2.85\pm0.10$&0.124&$ 94.2\pm31.6\pm11.7$\\
$K^{*0}K^-\pi^+\pi^0+c.c.$&$ 41.2\pm11.2$&$ 1.3\pm0.7$&$ 39.9\pm11.2$&$ 3.01\pm0.08$&0.172&$116.3\pm32.7\pm20.0$\\
$K^{*+}K^-\pi^+\pi^-+c.c.$&$ 22.3\pm 9.1$&$ 0.7\pm0.2$&$ 21.6\pm 9.1$&$ 2.18\pm0.07$&0.150&$173.9\pm73.3\pm26.1$\\
$\phi\pi^+\pi^-\pi^0$     & 2            &  0         &$<5.91$       &$ 3.11\pm0.06$&0.115&$<25.5$              \\
$\Lambda \bar\Lambda\pi^0$&0             &0           &$<2.44$       &$ 5.02\pm0.07$&0.123&$<7.9$               \\
 \hline
\end{tabular}
\label{tab:crs3773}
\end{center}
\end{table*}

\begin{table*}[hbtp]
\begin{center}
\caption{The observed cross sections for $e^+e^-\to f$ at
$\sqrt{s}=$ 3.650 GeV, where $N^{\rm obs}$ is the number of events
observed from the continuum data. The definitions of the other
symbols are the same as those in table \ref{tab:crs3773}.}
\begin{tabular}{|l|c|c|c|c|c|c|} \hline
\multicolumn{1}{|c|}{$e^+e^-\to$}&$N^{\rm obs}$&$N^{\rm b}$ &$N^{\rm net}$ (or $N^{\rm
up}$) &$\epsilon$[\%]&$\Delta_{\rm sys}$ &$\sigma$ (or
$\sigma^{\rm up}$) [pb] \\ \hline
$\omega\pi^+\pi^-$        & 4            &  0         &$<8.60$       &$3.34\pm0.09$&0.116&$<50.9$ \\
$\omega K^+K^-$           & 1            &  0         &$<4.36$       &$1.62\pm0.06$&0.119&$<53.4$ \\
$\omega p\bar p$          & 2            &  0         &$<5.91$       &$3.81\pm0.09$&0.124&$<30.9$ \\
$K^+K^-\rho^0\pi^0$       &$ 11.4\pm 6.4$&$2.3\pm0.9$ &$ 9.1\pm 6.4$ &$2.98\pm0.05$&0.224&$ 47.6\pm 33.4\pm10.7$\\
$K^+K^-\rho^+\pi^-+c.c.$  &$ 27.0\pm10.0$&$0.4\pm0.3$ &$26.6\pm10.0$ &$2.92\pm0.10$&0.139&$141.9\pm 53.3\pm19.7$\\
$K^{*0}K^-\pi^+\pi^0+c.c.$&$ 17.5\pm 7.9$&$0.5\pm0.4$ &$17.0\pm 7.9$ &$3.10\pm0.08$&0.140&$128.1\pm 59.5\pm17.9$\\
$K^{*+}K^-\pi^+\pi^-+c.c.$&$  9.7\pm 5.6$&$0.6\pm0.6$ &$ 9.1\pm 5.6$ &$2.25\pm0.07$&0.149&$189.0\pm116.3\pm28.2$\\
$\phi\pi^+\pi^-\pi^0$     & 2            &  0         &$<5.91$       &$3.17\pm0.06$&0.115&$<66.7$               \\
$\Lambda \bar\Lambda\pi^0$& 0            &  0         &$<2.44$       &$4.95\pm0.07$&0.123&$<21.4$               \\
\hline
\end{tabular}
\label{tab:crs3650}
\end{center}
\end{table*}

\subsection{Upper limits on the observed cross section and the
branching fraction for $\psi(3770)\to f$}

The upper limits $\sigma^{\rm up}_{\psi(3770)\to f}$ on the
observed cross sections for $\psi(3770)$ decay to the final states
$\omega\pi^+\pi^-$, $\omega K^+K^-$, $\omega p \bar p$,
$K^+K^-\rho^0\pi^0$, $\phi\pi^+\pi^-\pi^0$ and $\Lambda\bar
\Lambda\pi^0$ are directly set based on the upper limits on their
observed cross sections at 3.773 GeV.

For the other final states, the observed cross section for $\psi(3770)\to
f$ is determined by
\begin{equation}
\sigma_{\psi(3770)\to f}=
\sigma^{\rm 3.773\hspace{0.05cm}GeV}_{e^+e^-\to f} -
f_{\rm co}\times \sigma^{\rm 3.650\hspace{0.05cm}GeV}_{e^+e^- \to f},
\label{eq:obscrs}
\end{equation}
where $f_{\rm co}$ is the normalization factor in which we take
into account the 1/s dependence of the cross section and
neglect the difference in the corrections for the ISR and vaccum polarization
effects between the two energy points.
Inserting the values of $\sigma^{\rm
3.773\hspace{0.05cm}GeV}_{e^+e^- \to f}$ and $\sigma^{\rm 3.650
\hspace{0.05cm}GeV}_{e^+e^- \to f}$ listed in tables
\ref{tab:crs3773} and \ref{tab:crs3650}, and $f_{\rm co}$ in Eq.
(\ref{eq:obscrs}), we obtain the $\sigma_{\psi(3770)\to f}$ for
each mode, as shown in the second column of table
\ref{tab:up_psipp}, where the first error is the statistical, the
second is the independent systematic error (arising from the
uncertainties in the Monte Carlo statistics, in fitting to the
mass spectrum and in the background subtraction), and the third is
the common systematic error (arising from the other uncertainties
as discussed in the subsection B). The upper limits $\sigma^{\rm
up}_{\psi(3770)\to f}$ on the observed cross sections for
$\psi(3770)$ decay to these final states are set by shifting the
cross sections by $1.64\sigma$, where the $\sigma$ is the total
error of the measured cross section. We treat
$\sigma_{\psi(3770)\to f}$ with minus value as zero, and then set
its upper limit. The third column of table \ref{tab:up_psipp} shows
the results on $\sigma^{\rm up}_{\psi(3770) \to f}$.

The upper limit on the branching fraction for $\psi(3770)\to f$ is
set by
\begin{equation}
{\mathcal B}^{\rm up} _{\psi(3770)\to f}=\frac{\sigma^{\rm
up}_{\psi(3770) \to f}}{\sigma^{\rm obs}_{\psi(3770)}
\times \left [ 1-\Delta \sigma^{\rm obs}_{\psi(3770)} \right ]},
\label{eq:bfcrs}
\end{equation}
where $\sigma^{\rm obs}_{\psi(3770)}$ is the observed cross
section for the $\psi(3770)$ production, $\Delta \sigma^{\rm
obs}_{\psi(3770)}$ is the relative error in $\sigma^{\rm
obs}_{\psi(3770)}$. Here, $\sigma^{\rm obs}_{\psi(3770)}=
7.15\pm0.27\pm0.27$ nb \cite{crshads}, is obtained by weighting
two measurements \cite{brdd2,rval} from BES Collaboration.
Inserting the numbers of $\sigma^{\rm up}_{\psi(3770)\to f}$,
$\sigma^{\rm obs}_{\psi(3770)}$ and $\Delta \sigma^{\rm
obs}_{\psi(3770)}$ in Eq. (\ref{eq:bfcrs}), we obtain ${\mathcal
B}^{\rm up}_{\psi(3770) \to f}$ for each mode, as shown in the
fourth column of table \ref{tab:up_psipp}.

\newcommand{\rb}[1]{\raisebox{1.5ex}[0pt]{#1}}
\begin{table*}[htbp]
\begin{center}
\caption{ The upper limits on the observed cross section
$\sigma^{\rm up}_{\psi(3770)\to f}$ and the branching fraction
${\mathcal B}^{\rm up}_{\psi(3770)\to f}$ for $\psi(3770)\to f$
are set at 90\% C.L.. The $\sigma_{\psi(3770)\to f}$ in the second
column is calculated with Eq. (\ref{eq:obscrs}), where the first
error is the statistical, the second is the mode-dependent
systematic, and the third is the common systematic error. Here,
the upper $^t$ denotes that we treat the upper limit on the
observed cross section for $e^+e^-\to f$ at 3.773 GeV as
$\sigma^{\rm up}_{\psi(3770)\to f}$, the upper $^n$ denotes that
we neglect the contribution from the continuum production, and the
upper $^z$ denotes that we treat the central value of
$\sigma_{\psi(3770)\to f}$ as zero if it is less than zero. }
\begin{tabular}{|l|c|c|c|} \hline
   & $\sigma_{\psi(3770)\to f}$
   & $\sigma^{\rm up}_{\psi(3770)\to f}$
   & ${\mathcal B}^{\rm up}_{\psi(3770)\to f}$ \\
\multicolumn{1}{|c|}{\rb{Decay Mode}} & [pb] & [pb]    & [$\times 10^{-3}$] \\ \hline
$\omega\pi^+\pi^-$         &$<37.1^{tn}$                   & $<37.1$&$<5.5$  \\
$\omega K^+K^-$            &$<44.5^{tn}$                   & $<44.5$&$<6.6$  \\
$\omega p\bar p$           &$<20.3^{tn}$                   & $<20.3$&$<3.0$  \\
$K^+K^-\rho^0\pi^0$        &$<5.6^{tn}$                    &$<5.6$  &$< 0.8$ \\
$K^+K^-\rho^+\pi^-+c.c.$   &$-38.6\pm 59.0\pm11.9\pm 4.4^z$&$<99.0$ &$<14.6$ \\
$K^{*0}K^-\pi^+\pi^0+c.c.$ &$ -3.6\pm 64.6\pm18.2\pm 0.4^z$&$<110.0$&$<16.2$ \\
$K^{*+}K^-\pi^+\pi^-+c.c.$ &$ -3.0\pm131.2\pm24.5\pm 0.3^z$&$<218.9$&$<32.3$ \\
$\phi\pi^+\pi^-\pi^0$      &$<25.5^{tn}$                   &$<25.5$ &$<3.8$  \\
$\Lambda \bar\Lambda\pi^0$ &$<7.9^{tn}$                    &$<7.9$  &$<1.2$  \\
\hline
\end{tabular}
\label{tab:up_psipp}
\end{center}
\end{table*}

\section{Summary}
In summary, by analyzing the data sets taken at $\sqrt{s}=$ 3.773
and 3.650 GeV with the BES-II detector at the BEPC collider, we
have measured the observed cross sections for
$\omega \pi^+\pi^-$, $\omega K^+K^-$, $\omega p\bar p$,
$K^+K^-\rho^0\pi^0$, $K^+K^-\rho^+\pi^-+c.c.$,
$K^{*0}K^-\pi^+\pi^0+c.c.$, $K^{*+}K^-\pi^+\pi^-+c.c.$,
$\phi \pi^+\pi^-\pi^0$ and $\Lambda \bar
\Lambda \pi^0$ produced in $e^+e^-$ annihilation at the two energy
points.
Upper limits with 90\% confidence level were derived for the observed cross sections and
branching fractions for $\psi(3770)$ decay to these final states.
We do not observe significant difference between the observed cross
sections for most exclusive light hadron final states at the two energy points.
However, this does not mean that $\psi(3770)$ does not decay into these final states,
since we neglect the interference effects between the
continuum and resonance amplitudes. A better way to extract the
branching fractions for $\psi(3770)\to$ {\it exclusive light hadrons}
would be to analyze their energy-dependent observed cross sections
at more energy points covering both $\psi(3770)$ and $\psi(3686)$
\cite{crsscan} with the coming BES-III detector at the BEPC-II
collider. However, the observed cross sections reported in this
paper and those reported in Refs. \cite{crshads,crshads2,huang,adams} provide
constraints which could help to understand both the mechanism
of the continuum light hadron production and the discrepancy
between the observed cross sections for $D\bar D$ and $\psi(3770)$
production.

\section{acknowledgments}
The BES collaboration thanks the staff of BEPC for their hard
efforts. This work is supported in part by the National Natural
Science Foundation of China under contracts Nos. 10491300,
10225524, 10225525, 10425523, the Chinese Academy of Sciences
under contract No. KJ 95T-03, the 100 Talents Program of CAS under
Contract Nos. U-11, U-24, U-25, the Knowledge Innovation Project
of CAS under Contract Nos. U-602, U-34 (IHEP), the National
Natural Science Foundation of China under Contract  No. 10225522
(Tsinghua University).

\end{document}